\documentclass[a4paper]{jpconf}
\usepackage{graphicx,amsmath}

\newcommand{\amc}{{\sc Mad\-Graph5\textunderscore}a{\sc MC@NLO}}
\def\be{\begin{equation}}
\def\ee{\end{equation}}
\newcommand\sss{\scriptscriptstyle}
\def\bsp#1\esp{\begin{split}#1\end{split}}

\begin{document}
\title{Opportunities with top quarks at \\ future circular colliders}

\author{Benjamin Fuks}

\address{CERN, PH-TH, CH-1211 Geneva 23, Switzerland}
\address{Institut Pluridisciplinaire Hubert Curien/D\'epartement
    Recherches Subatomiques, Universit\'e de Strasbourg/CNRS-IN2P3,
    23 rue du Loess, F-67037 Strasbourg, France}

\ead{benjamin.fuks@iphc.cnrs.fr}

\begin{abstract}
We describe various studies relevant for top physics at
future circular collider projects currently under discussion. We show
how highly-massive top-antitop systems produced in proton-proton collisions
at a center-of-mass energy of 100~TeV could be observed and employed for
constraining top dipole moments, investigate the reach of future proton-proton
and electron-positron machines to
top flavor-changing neutral interactions, and discuss top parton densities.
\end{abstract}

\section{A future circular collider facility at CERN}
The Large Hadron Collider (LHC) at CERN has delivered very high quality results
during its first run in 2009-2013, with in particular the discovery
of a Higgs boson with a mass of about 125~GeV in
2012. Unfortunately, no hint for the
presence of particles beyond the Standard Model has been
observed. Deviations from the Standard Model are however still allowed and
expected to show up either
through precision measurements of indirect probes, or directly at collider
experiments. In this context, high precision would require to push the
intensity frontier further and further, whereas bringing the energy frontier
beyond the LHC regime would provide handles on new kinematical thresholds.
Along these lines, a design study for a new accelerator facility aimed to
operate at CERN in the post-LHC era has been undertaken. This study
focuses on a machine that could collide protons at a center-of-mass energy of
$\sqrt{s}=100$~TeV, that could be built in a tunnel of about 80-100~km in the
area of Geneva and benefit from the existing infrastructure at
CERN~\cite{fcchh}.
A possible intermediate step in this project
could include an electron-positron machine with a collision
center-of-mass energy ranging from 90~GeV (the $Z$-pole) to 350~GeV (the
top-antitop
threshold), with additional working points at $\sqrt{s}=160$~GeV (the $W$-boson
pair production threshold) and 240~GeV (a Higgs factory)~\cite{fccee}. In parallel,
highly-energetic lepton-hadron and heavy ion collisions are also under
investigation.

Both the above-mentioned future circular collider (FCC) setups are expected to
deliver a copious amount of top quarks. More precisely, one trillion of them
are expected to be produced in 10~ab$^{-1}$ of proton-proton collisions at
$\sqrt{s}=100$~TeV and five millions of them in the same amount of
electron-positron collisions at $\sqrt{s}=350$~GeV (which will in particular
allows for top mass and width measurements at an accuracy of about
10~MeV~\cite{Gomez-Ceballos:2013zzn}). This consequently opens
the door for an exploration of the properties of the top quark, widely
considered as a sensitive probe to new physics given its mass close to the
electroweak scale, with an unprecedented accuracy.
This is illustrated below with three selected examples.

\section{Top pair production in 100~TeV proton-proton collisions}
The top quark pair-production cross section for proton-proton collisions at
$\sqrt{s}=100$~TeV reaches 29.4~nb at the next-to-leading order accuracy in QCD,
as calculated with \amc~\cite{Alwall:2014hca} and the NNPDF 2.3 set of parton
densities~\cite{Ball:2012cx}. A very large number of $t\bar t$
events are thus expected to be produced for integrated luminosities of several
ab$^{-1}$, with a significant number of them featuring a top-antitop system
whose invariant-mass lies in the
multi-TeV range. Whereas kinematical regimes never probed up to now will become
accessible, standard $t\bar t$ reconstruction techniques may not be sufficient
to observe such top quarks that are highly boosted, with a
transverse momentum ($p_T$) easily exceeding a few TeV. In addition, it is not
clear how current boosted top tagging techniques, developed in the context of
the LHC, could be applied. Consequently, it could be complicated to distinguish
a signal made of a pair of highly boosted top quarks from the overwhelming
multijet background.

\begin{figure}[t]
\centering
\includegraphics[width=.57\columnwidth]{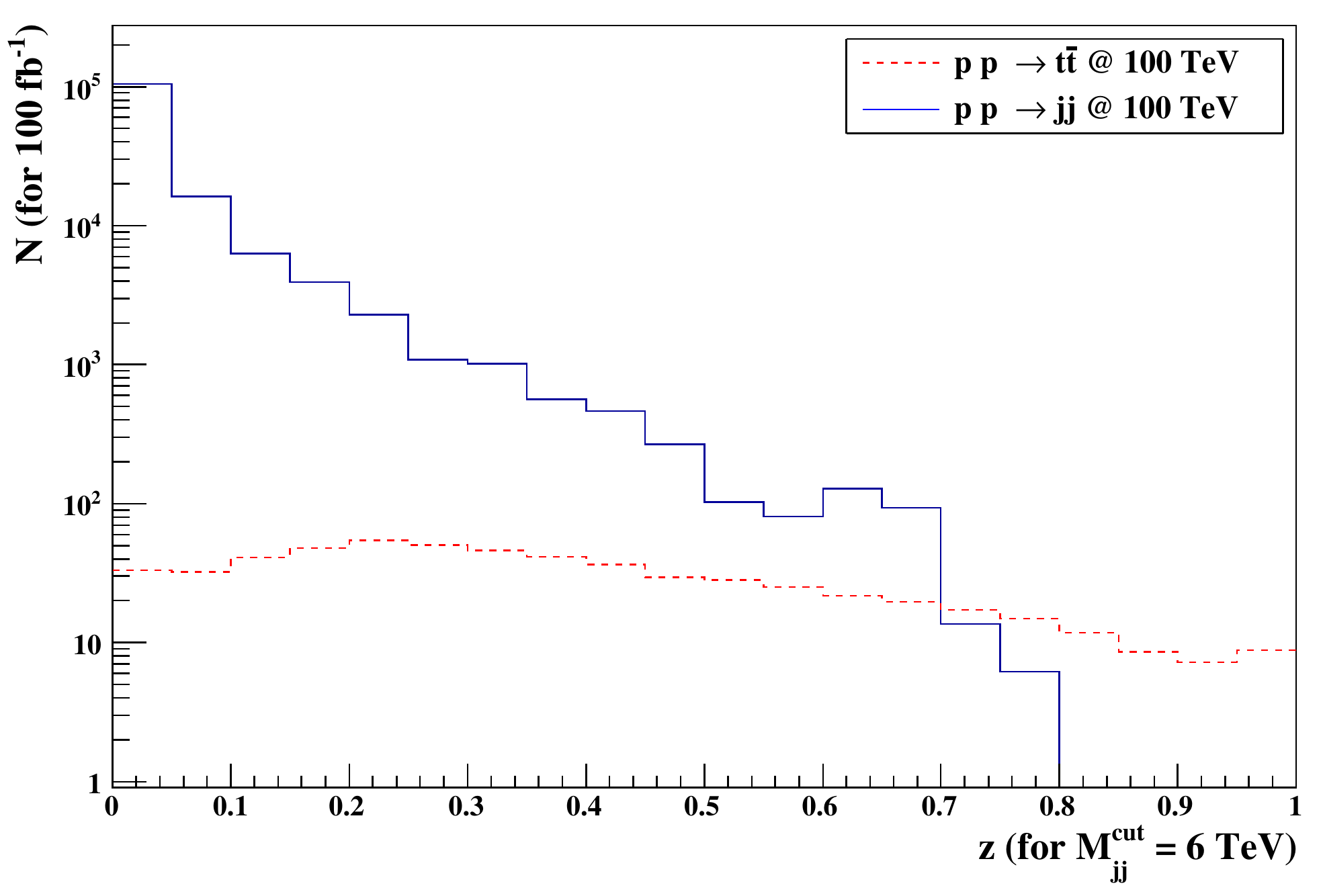}
\includegraphics[width=.41\columnwidth]{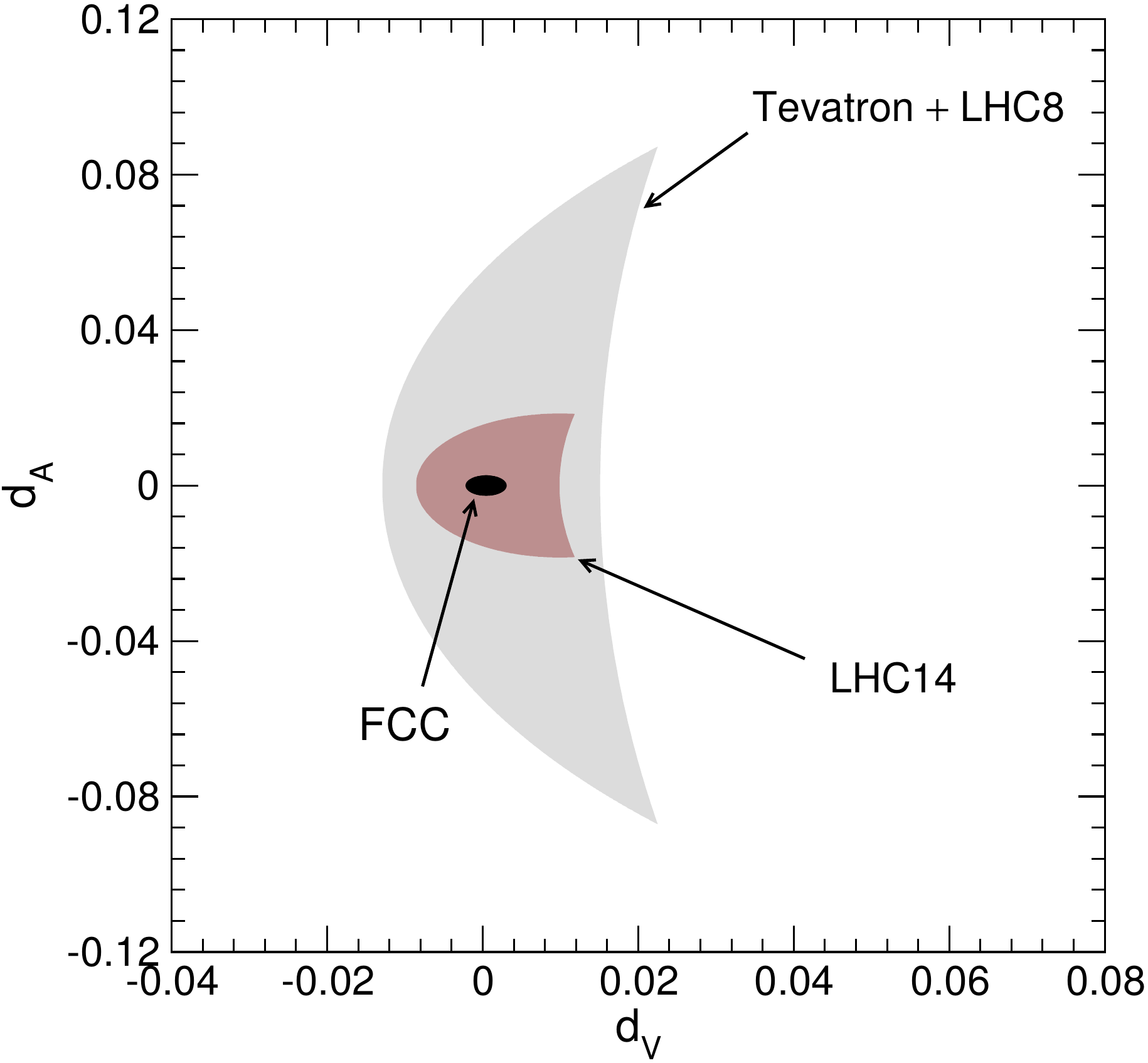}
\caption{\label{fig:ttbar}\small \textit{Left}: distributions of the $z$ variable
  of Eq.~(\ref{eq:z}) for proton-proton collisions at $\sqrt{s}=100$~TeV. We
  present predictions for top-antitop (red dashed) and multijet
  (blue plain) production, after selecting events as described in the text. We have
  fixed $M_{jj}^{\rm cut}$ to 6~TeV and normalized the results to
  100~fb$^{-1}$. \textit{Right}: constraints on the top dipole moments
  derived from measurements at the Tevatron and the LHC (gray), and from
  predictions at the LHC (red, $\sqrt{s}=14$~TeV) and the FCC (black,
  $M_{jj}^{\rm cut} = 10$~TeV).}
\end{figure}

To demonstrate that this task is already manageable with basic
considerations~\cite{inprep}, we have analyzed, by means
of the {\sc MadAnalysis}~5 package~\cite{Conte:2012fm}, leading-order
hard-scattering events simulated with \amc\ and matched to the
parton showering and hadronization
algorithms included in {\sc Pythia}~8~\cite{Sjostrand:2007gs}. We
have considered, in our analysis, jets with a $p_T > 1$~TeV that have
been reconstructed
with {\sc FastJet}~\cite{Cacciari:2011ma} and an anti-$k_T$ jet
algorithm with a radius parameter $R=0.2$~\cite{Cacciari:2008gp}.
We preselect events featuring at least two jets with a
pseudo\-ra\-pi\-di\-ty $|\eta|<2$ and at least one muon lying in a cone of
$R=0.2$ of any of the selected jets.
The invariant mass of the system comprised of the two leading jets
is additionally constrained to be larger than
a threshold $M_{jj}^{\rm cut}$. We then investigate the properties of the
selected muons relatively to those of the related jet.
In this context, we present
on Figure~\ref{fig:ttbar} (left) the distribution in a $z$ variable
defined as the ratio of the muon transverse momentum $p_T(\mu_i)$ to the
corresponding jet transverse momentum $p_T(j_i)$, maximized over the $n$
final-state muons of the event,
\be
  z \equiv \max_{i=1,\ldots n} \frac{p_T(\mu_i)}{p_T(j_i)} \ .
\label{eq:z}\ee
Muons arising from multijet events are mostly found to carry a small
fraction of the jet transverse momentum, which is inferred by their production
mechanism ($B$- and $D$-meson decays). This contrasts with muons induced
by prompt decays of top quarks that can gather a significant fraction of the top
$p_T$. Imposing the $z$-variable to be larger than an optimized threshold
$z^{\rm cut}$, it becomes possible to obtain signal over background ratios $S/B$
of order one and
extract the $t\bar t$ signal at the $5\sigma$ level (defined by $S/\sqrt{S+B}$).
We study, in Table~\ref{tab:ttbar}, the $z^{\rm cut}$ value for different
invariant-mass threshold $M_{jj}^{\rm cut}$, and present
the associated $S/B$ ratio together with the
luminosity necessary for a signal extraction at $5\sigma$.

\begin{table}
\caption{\label{tab:ttbar}\small Values of the  $z^{\rm cut}$ parameter
  for different $M_{jj}^{\rm cut}$ choices. We also present 
  the corresponding $S/B$ ratio and the luminosity ${\cal L}_{5\sigma}$
  necessary for a $5\sigma$ extraction of a $t\bar t$ signal from the multijet
  background.}
\begin{center}
\begin{tabular}{llll}
\br
  $M_{jj}^{\rm cut}$ & $z^{\rm cut}$ & $S/B$ & ${\cal L}_{5\sigma}$\\
\mr
6~TeV  & 0.5 & 0.39 & 36.1 fb$^{-1}$\\
10~TeV & 0.5 & 0.74 & 202  fb$^{-1}$\\
15~TeV & 0.4 & 0.25 & 2.35 ab$^{-1}$\\
\br
\end{tabular}
\end{center}
\end{table}

On Figure~\ref{fig:ttbar} (right), we illustrate how a measurement of the
fiducial cross section related to the above selection with
$M_{jj}^{\rm cut} = 10$~TeV
could be used to constrain the
top chromomagnetic and chromoelectric dipole moments $d_V$ and $d_A$. In our
conventions, they are defined by
\be
  \mathcal{L} = \frac{g_s}{m_t} \bar t \sigma^{\mu \nu}
    (d_V + i d_A \gamma_5) T_a t \, G_{\mu \nu}^a \ ,
\ee
where $g_s$ denotes the strong coupling, $T^a$ the fundamental representation
matrices of $SU(3)$, $m_t$ the top mass and
$G_{\mu\nu}$ the gluon field strength tensor. We have imported this Lagrangian
into \amc\ by using {\sc FeynRules}~\cite{Alloul:2013bka,Degrande:2011ua} to
estimate the new
physics contributions to the $t\bar t$ signal and extract bounds on the top
dipole moments. While current limits have been derived from total
rate measurements at the Tevatron and the LHC (gray), the FCC predictions
with $M_{jj}^{\rm cut} = 10$~TeV
(black) correspond to 1~ab$^{-1}$ of collisions, and we have superimposed
the expectation for the future LHC run at
$\sqrt{s} = 14$~TeV (red) after setting $M_{jj}^{\rm cut} = 2$~TeV and using
standard top tagging efficiencies~\cite{CMS:2014fya}. The FCC is hence
expected to constrain top dipole moments
to lie in the ranges
$-0.0022 < d_V < 0.0031$ and $|d_A| < 0.0026$, which gets close
to the reach of indirect probes like the electric dipole
moment of the neutron ($|d_A|<0.0012$~\cite{Kamenik:2011dk}) or
$b\to s \gamma$ transitions ($-0.0038 < d_V < 0.0012$~\cite{Martinez:2001qs}).

\section{Flavor-changing neutral interactions of the top quark}
In the Standard Model, the top flavor-changing couplings to the
neutral bosons are suppressed due the unbroken QCD and QED symmetries
and the GIM mechanism. Many new physics extensions however predict an
enhancement of those interactions, whose hints are therefore
searched for either in the anomalous
decay of a $t\bar t$ pair, or in an anomalous single top
production. An effective approach for describing those effects
consists of supplementing the Standard Model Lagrangian by
dimension-six operators that give rise
to a basis of top anomalous couplings that can be chosen
minimal~\cite{AguilarSaavedra:2008zc}. Taking the example of the three
operators
\be
  {\cal L} =
    \frac{\bar c_{\sss H}}{\Lambda^2} \Phi^\dag \Phi\ \Phi^\dag\cdot{\bar Q}_L u_R
  + \frac{\bar c_{\sss W}}{\Lambda^2} \Phi^\dag \cdot \big({\bar Q}_L T_{2k}\big)
        \sigma^{\mu\nu} u_R\ W_{\mu\nu}^k
  + \frac{\bar c_{\sss G}}{\Lambda^2} \Phi^\dag \cdot {\bar Q}_L \sigma^{\mu\nu} T_a u_R G_{\mu\nu}^a
 + {\rm h.c.}\ ,
\label{eq:l6}\ee
where flavor indices are understood for clarity, we indeed observe that
flavor-changing top couplings to the Higgs boson, the $Z$-boson and
the gluon are induced after electroweak symmetry breaking. In our
notation, $\Lambda$ denotes the new physics scale, $\Phi$ ($Q$) a weak
doublet of Higgs (left-handed quark) fields and $u_R$ a right-handed up-type
quark field. Assuming the Wilson coefficients $\bar c$
of order 1, current ATLAS and CMS data constrains $\Lambda\simeq$
1~TeV (4-5 TeV in the case of the ${\cal O}_{\sss G}$
operator)~\cite{topatlas,b2gcms}. A naive estimate of the FCC sensitivity to
$\Lambda$ can then be derived from these numbers by rescaling both the signal and
background with the
relevant cross sections and luminosities. Assuming 10~ab$^{-1}$ of 100~TeV
proton-proton collisions, one finds that the LHC limits could be increased
by a factor of about 20~\cite{zupan}.

Limits on new physics operators such as those of
Eq.~(\ref{eq:l6}) could also be obtained in electron-positron collisions at high
energies by relying, for instance, on single top
production~\cite{Khanpour:2014xla}. Signal and background have been
simulated using the tools introduced in the previous section
together with a modeling of the detector effects by Gaussian smearing.
Exploitation of the kinematical configuration of the signal
with a multivariate technique has then been found
sufficient to extract the signal from the background and
derive bounds on the anomalous top interactions, that have
been translated in terms of limits on rare top decay
branching ratios on Figure~\ref{fig:fcncpdf} (left).

\section{Top parton density in the proton}

In proton-proton collisions at $\sqrt{s}=100$~TeV, all quarks including the
top would appear essentially massless, so that it may seem appropriate to
investigate processes with initial-state top quarks. While a
six-flavor-number scheme (6FNS) allows for the resummation of collinear
logarithms of the process scale over the top mass into a top density, this
is only justified when these logarithms are large, the
five-flavor-number scheme (5FNS) calculation spoiling in this case perturbative QCD.
Both the 5FNS and 6FNS computations
can however be consistently matched to guarantee accurate predictions for
any scale. In the ACOT
scheme~\cite{Aivazis:1993pi}, the 5FNS and 6FNS results are summed and the
matching is achieved by subtracting from the top density $f_t$ its leading
logarithmic approximation $f_t^0$ that is
already included in the 5FNS calculation,
\be\bsp
  \sigma(pp\to X^0) =&\
    \Big[f_t - f_t^0\Big] \otimes  \Big[f_{\bar t} - f_{\bar t}^0\Big]
       \otimes \sigma(t\bar t \to X^0)
    + \Big[f_t - f_t^0\Big] \otimes  f_g
       \otimes \sigma(tg \to X^0 t) \\
 &\ + f_g\otimes \Big[f_{\bar t} - f_{\bar t}^0\Big]
       \otimes \sigma(g\bar t \to X^0 \bar t)
    + f_g\otimes f_g
       \otimes \sigma(gg \to X^0 t\bar t) \ ,
\esp\label{eq:acot}\ee
where $X^0$ denotes any electrically and color neutral final state and $f_g$ the
gluon density.

\begin{figure}[t]
\centering
\includegraphics[width=.45\columnwidth]{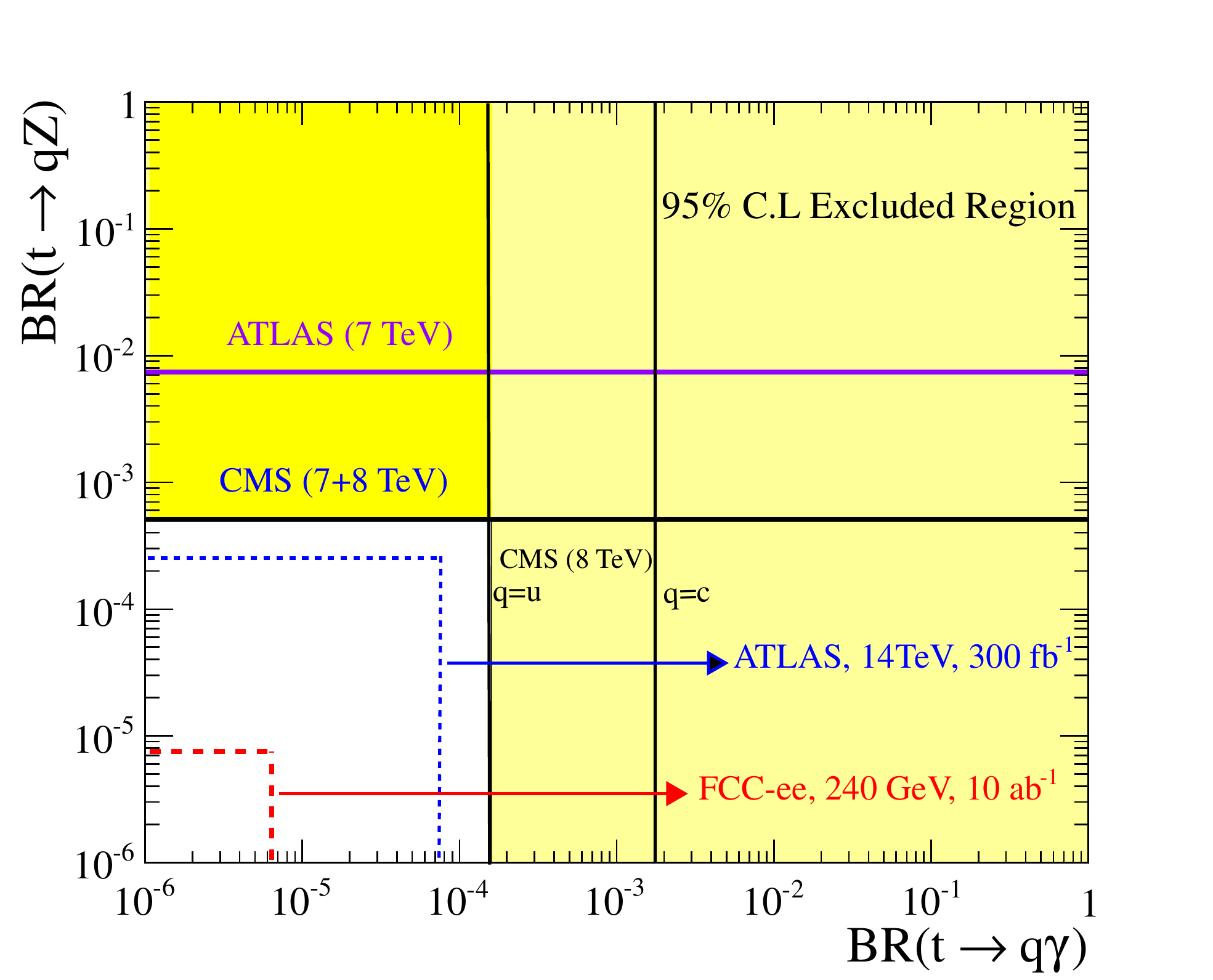}
\includegraphics[width=.53\columnwidth]{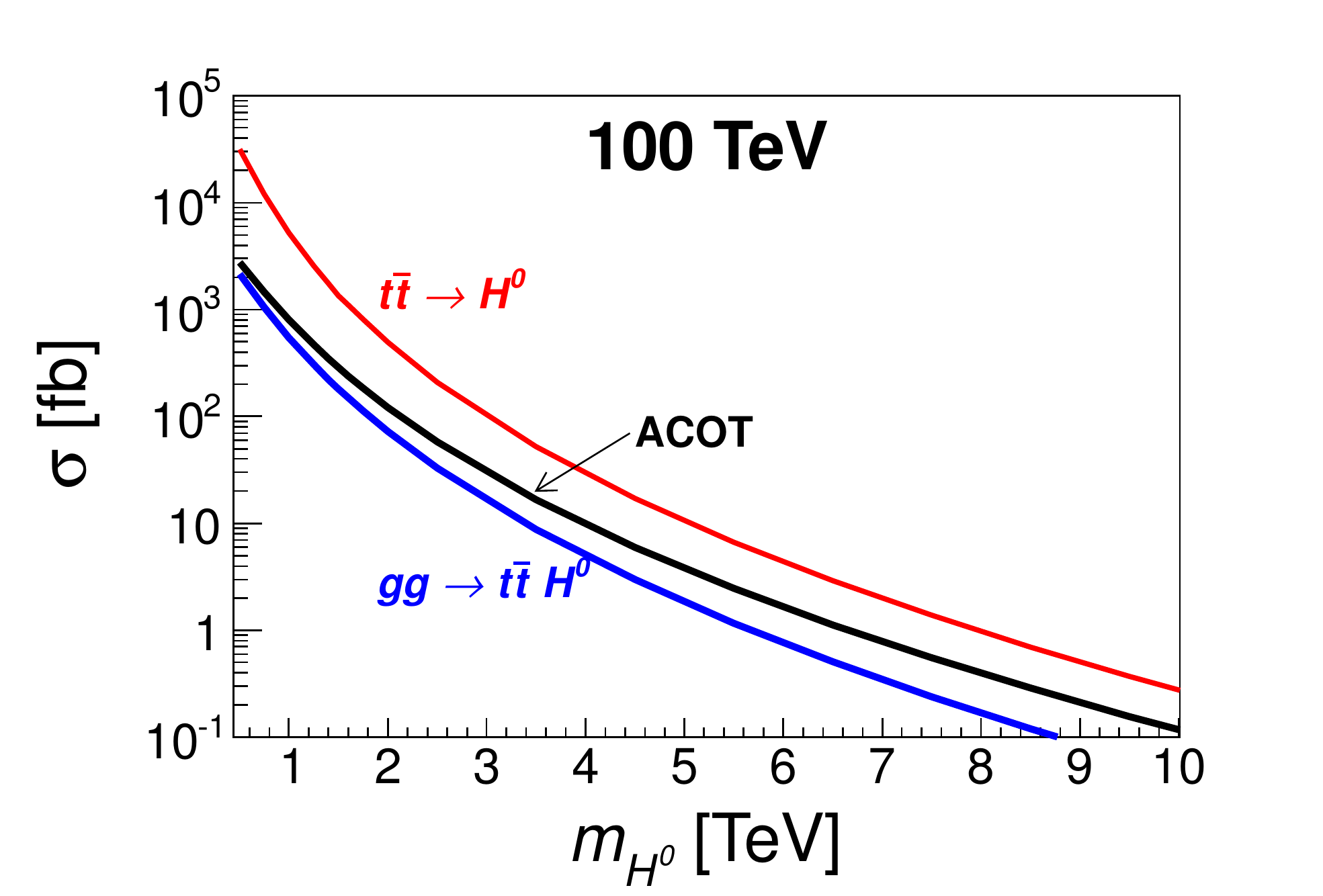}
\caption{\label{fig:fcncpdf}\small \textit{Left}: limits on top decays into
  a neutral electroweak boson and a lighter quark.
  Current LHC bounds have been indicated, together with the expectation
  for 10~ab$^{-1}$ of electron-positron collisions at $\sqrt{s}=240$~GeV.
  Figure taken from Ref.~\cite{Khanpour:2014xla}. \textit{Right}:
  Total cross section for the production of
  a heavy Higgs boson in the 6FNS (red), 5FNS (blue) and ACOT scheme (black)
  for proton-proton collisions at $\sqrt{s}=100$~TeV. Figure taken from
  Ref.~\cite{Han:2014nja}.}
\end{figure}

Figure~\ref{fig:fcncpdf} (right) shows the production of a heavy neutral Higgs
boson~\cite{Han:2014nja}, and compare leading-order
predictions in the 5FNS (blue), 6FNS (red)
and ACOT scheme (black) for proton-proton collisions at
$\sqrt{s}=100$~TeV. For small Higgs
masses, the subtraction of the leading logarithmic terms in Eq.~(\ref{eq:acot})
cancels almost entirely the 6FNS contribution, the ACOT result mostly being
the 5FNS ones. In this region, the logarithms in
the top mass are small so that their resummation into a top density is not
justified. For larger masses, they start to
play a role, although the use of the 6FNS alone still yields a
large overestimation of the cross section. Predictions including top densities
should consequently be matched to the 5FNS result, as also found
for charged Higgs production~\cite{Dawson:2014pea}, and not employed
as such.

\section{Conclusions}
We have discussed three top physics cases that are
relevant for collisions at future circular colliders. We have shown how highly
massive top-antitop systems could be observed in pro\-ton-\-pro\-ton collisions at a
center-of-mass energy of 100~TeV and further used to constrain top dipole
moments. We have then sketched how constraints on top flavor-changing neutral
interactions would improve both at future proton-proton and electron-positron
colliders, and finally investigated the issue of the top parton
density, relevant for proton-proton collisions at energies much
larger than the top mass.

\ack
I am grateful the conference organizers for setting up this very nice Top2014 event,
as well as to J.A.~Aguilar-Saavedra, M.L.~Mangano and T.~Melia for
enlightening discussions on top physics at the FCC. I am also grateful
to T.~Han,
H.~Khanpour, S.~Khatibi, M.~Khatiri Yanehsari, M.M.~Najafabadi,
J.~Sayre and S.~Westhoff for providing material included in this report.
This work has been partly supported by the LHC Physics Center at CERN (LPCC).

\section*{References}
\bibliographystyle{iopart-num}
\bibliography{fuks}

\end{document}